# Double-Ionization Satellites in the X-ray Emission Spectrum of Ni Metal

Ryan A. Valenza, Evan P. Jahrman, Joshua J. Kas, and Gerald T. Seidler(*)

Physics Department, University of Washington, Seattle WA 98195-1560

We report measurements of the nonresonant x-ray emission spectroscopy (XES) from Ni metal in an energy range spanning the K$\beta$, valence-to-core, and double-ionization (DI) satellites that appear beyond the single-particle Fermi level. We make special use of a laboratory-based x-ray spectrometer capable of both x-ray emission and x-ray absorption measurements to accurately align the XES and x-ray absorption spectra to a common energy scale. The careful alignment of energy scales is requisite for correction of the strong sample absorption of DI fluorescence above the Ni K-edge energy. The successful correction of absorption effects allows a determination of the branching ratios for the [1s3d], [1s3p], [1s2p] and [1s2s] satellites with respect to their corresponding diagram lines. We compare our results with other work, finding good agreement with prior experiments and with theoretical calculations in the multi-configuration Dirac-Fock framework.

I.   **Introduction**

Multi-electron transitions caused by the absorption of single x-ray photons were first observed a century ago by Siegbahn and Stenstrom[1] before further study by researchers including Richtmyer[2, 3] and Druyvesteyn[4]. Since then, a range of phenomena have been attributed to these processes, including numerous features in nonresonant x-ray emission spectroscopy (XES)[5-7], low energy satellites in x-ray photoelectron spectroscopy (XPS)[8-11], and discontinuities in x-ray absorption fine structure spectra[12].

The simplest model for multi-electron excitations employs the shake process within the sudden approximation. Within this model, the incident photon first induces the emission of a single photoelectron, causing all remaining electrons to subsequently experience a change in the central potential due to the reduction in screening accompanying the creation of a core hole. Second, the reduction in screening prompts the occupied orbitals to relax, yielding an imperfect overlap between the initial and final wavefunctions. Finally, this perturbation results in a nonzero probability that a second electron will undergo a monopole excitation to an unoccupied bound state (a shake-up process, 'SU') or to a continuum state (a shake-off process, 'SO'). An extensive record of theoretical studies of shake probabilities in the sudden approximation[13-17] was motivated by the 1941 observation of Migdal and Feinberg[18, 19] that the rapid change in nuclear potential following β decay results in an appreciable probability of ionization in each of the atom's occupied orbitals. Recently, *ab initio* relativistic Dirac-Fock multiplet calculations including configurations with spectator holes arising from shake processes have enabled the accurate reconstruction of the emission spectra of Cu, Sc, and Ti[20-22]. The best agreement with experiment is achieved in the multi-configuration framework advocated by Chantler, Lowe, and Grant[23-25], implementing the procedure for transition probability calculations using nonorthogonal orbitals developed by Olsen *et al*[26].

Alternative models have been proposed to explain spectral features without the inclusion of shake effects. Of particular note are conduction-band collective excitations[27], exchange[28], surface plasmons[29], and the (e,2e)-like electron impact half collision knockout (KO) process[30]. As might be expected, progress toward understanding these spectral features has been incrementally provided by numerous experimental and theoretical studies. It is now known that the KO process, while measurable in many studies, becomes negligible at high energy excitations where photoabsorption approaches the asymptotic limit of shakeoff[31, 32], as is the case in this work. Also, the surface plasmon hypothesis has been called into question by Karis *et al*[33] in a careful XPS study of metallic Ni.

Irrespective of the microscopic description, double ionization (DI) is by far the most probable multi-electron transition[16] in typical experiments and, as a result, is the most commonly studied. More recently, a greater understanding of DI transitions has motivated several novel research directions including the emergence of the spin-flip forbidden $K\alpha_1^h$ ($^3P_1 \to {}^1S_0$) transition as a highly sensitive indicator of the transition from the LS coupling scheme to intermediacy[30, 34], experiments probing the variation of fundamental constants in space-time[35], and tests of the Breit interaction in quantum electrodynamics[36-38].

That said, much of the interest in multi-electron transitions stems from the observation that for weak interactions, such as those involving photoionization, the probability of DI events greatly exceeds predictions that treat both electrons as independent[39]. Consequently, the ejection of multiple electrons depends strongly on many-electron interactions[40] and can thus provide a means to understand intra-atomic electron correlations and verify theoretical methods as they are developed.

It is important to note that the DI process is not restricted to the high energy isothermal region, but is also present in the adiabatic regime close to the double-ionization threshold. In this limit, the potential of the first photoelectron during the second ionization cannot be ignored and is addressed, for example, via time-dependent perturbation theory[41, 42]. The adiabatic regime, and especially the transition from adiabatic to isothermal regime, has benefitted from many outstanding experimental efforts[34, 43, 44]. In such studies, satellite intensities at various excitation energies are fit by optimizing the parameters required by the Thomas[41], Roy[45], Roy-2[46], or Vatai[47] model, with the Thomas model the most common of the four. Nevertheless, it should be noted that these models are not *ab inito* treatments[48]. For example, only the Vatai and Roy model incorporate the Coulomb interaction as the mechanism of excitation while the Thomas and Roy-2 models simply postulate that the time dependence is described by an analytic function and include the interaction Hamiltonian in a parameter representing the asymptotic value given by the sudden approximation.

In the present study, we investigate the DI x-ray emission satellites occurring above the single-electron Fermi level of Ni in the isothermal limit of high-energy excitation. The various DI XES peaks observed above the Fermi level are analyzed with an eye toward establishing a protocol for reliable determination of the branch ratio of DI features to their corresponding diagram line fluorescence. These branch ratios serve as a natural benchmark for comparison to theory. In addition, several aspects of our experimental approach may prove useful in the future, particularly a method to align XES and XAFS spectra to a common energy scale when using a laboratory-based spectrometer and a demonstration of the necessary corrections for the sample's internal absorption that otherwise alter the intensity of DI XES features appearing above the single-particle Fermi level.

This manuscript continues as follows. First, in Section II, we describe the experimental details. This includes a description of instrumentation and data collection protocol, a detailed discussion of the energy-dependent correction for self-attenuation effects, and our method for obtaining absorption and emission spectra on the same energy scale. We argue that these latter two issues are critical for obtaining accurate estimates of the relative branch ratios of multi-electron satellites above the single-particle Fermi level. Next, in Section III, we present and discuss the results of the study. We give special attention to the energies and branch ratios of the several observed DI features, and their comparison with theory. Finally, in Section IV, we conclude.

## II. Experimental Procedure

### A. Laboratory spectrometer

Our group has recently developed laboratory-based (i.e., non-synchrotron) instrumentation for XES and x-ray absorption fine structure (XAFS) measurements[49-52]. Energy scanning, whether for XAFS or for XES, uses a scissors-style monochromator that symmetrically moves the source and detector while maintaining the delicate angular-orientation of the spherically bent crystal analyzer (SBCA) needed for alignment. The careful use of internal shielding together with the added rejection provided by the energy-dispersing silicon drift detector (Amptek SDD-123) results in exceptionally low backgrounds, allowing for clear observation of the DI features without use of any background subtraction in the measured XES spectra.

In the XAFS implementation of the instrument used here[49], photons from the x-ray tube source are monochromatized and refocused at the detector by a synchrotron-quality SBCA, providing useful flux for transmission-mode studies with 1-eV or finer energy resolution. The XES implementation benefits from the same inherent high resolution, but differs in that nonresonant excitation of a material is accomplished by direct illumination of the sample behind an entrance slit. The entrance slit, then, establishes an effective source on the Rowland circle and stabilizes instrument performance[52]. Additionally, direct illumination by the x-ray tube, whose output spectrum includes bremsstrahlung and characteristic fluorescence lines, is a highly efficient source of excitation as all photons above the Fermi level can create a core hole in the sample. While the x-ray tube (Moxtek Au anode) has 10 W maximum electron beam power, the close approach of the sample to the anode results in a core-hole generation rate that is intermediate between those of a monochromatized bending magnet and monochromatized insertion device beamline at a 3$^{rd}$ generation synchrotron[49, 52].

### B. Implementation of a common energy scale for emission and absorption

In the section below (II.C), we describe a method to correct for the strong absorption of the above-Fermi level DI emission by the sample itself, *i.e.*, because this x-ray emission has intensity at and above the absorption edge. This method requires, however, that the XES and XANES spectra be reliably placed onto the same energy scale; otherwise the steep rise in absorption at the edge will not be properly located and result in systematic error. Fortunately, we can make use of a novel feature of our laboratory-based instrument; its ability to transition simply between XES and XANES measurements[49]. Typically, this transition involves a reconfiguration of components along the Rowland circle, which has the possibility of introducing a shift in energy scale due to imperfect rigidity of the support structure for the source stage and especially the spectrometer entrance slit. We address this issue with a simple, localized procedure for energy alignment across the various measurements needed to correct XES spectra for the phenomenon of self-attenuation.

In Fig. 1 we demonstrate our procedure to obtain a XANES and XES spectrum on the same energy scale. In each subfigure, the SBCA is used for energy selection and focusing, and energy scanning is done by moving both the sample and detector symmetrically around the Rowland circle. First, in Fig. 1a, we show a typical XES geometry where the entire spectrum of the x-ray tube source illuminates the sample, causing it to fluoresce. Next, we place the sample in front of the detector and move the source onto the Rowland circle, as shown in Fig. 1b. for the XANES configuration. Here, we measure transmission through the sample to obtain the absorption cross section, μ, via the Beer-Lambert Law. The standard procedure is to then set the global energy scale by aligning the measured absorption edge to database values or XANES spectra recorded at a beamline or available in the HEPHAESTUS software package[53]. However, the finite rigidity of the spectrometer means that the reconfiguration of components can lead to a shift in energy scale between the XES and the XANES measurements.

Finally, an intermediate, hybrid configuration (Fig. 1c) helps resolve this difficulty by moving the sample to the opposite side of the Rowland circle so that it is in front of the source, but behind the entrance slit. As all components except the sample have been held fixed, we have high confidence that our energy scale has remained unaltered between the XAFS and hybrid configurations. The hybrid-configuration spectrum contains both an absorption edge (as we are measuring transmission through the sample) and emission peaks (as the sample is on the source side of the Rowland circle and is strongly excited by the x-ray tube in this geometry). The absorption edge can be aligned with the previously measured XANES spectrum, which is set to a global energy scale, and the observed fluorescence peaks in the hybrid spectrum can be used to shift the XES spectrum from the configuration of Fig. 1a onto the necessary common energy scale.

### C. Correction for combined geometric and absorptive effects

In this subsection, we describe the correction for geometric and absorptive effects to the intensity of the above Fermi level DI XES relative to the usual single-excitation diagram lines. These absorptive effects are known to influence the shape of spectral features in XES[54]. Indeed, other authors have taken steps to correct for sample absorption, particularly when deemed as requisite for reporting a quantitative result, such as for measurements of the magnetic circular dichroism of Gd films measured via XES[55]. In the present case, the intensity of spectral features above the Fermi level is strongly suppressed by absorptive effects.

In Fig. 2, we show a typical sample geometry for XES. Incoming photons from the tube source travel a distance $z/\sin(\alpha)$ before being absorbed by the sample, causing the emission of photons of energy $E_e$ to fill the core-hole. These emitted photons can then be reabsorbed by the sample during their exit path over the distance $z/\sin(\beta)$. When the parameter $z$ is integrated over the sample thickness $t$, we obtain the form of the measured spectrum, $I(E_e)$,

$$I(E_e) = \frac{\Omega}{4\pi} I_o(E_i) \epsilon(E_e|E_i) \frac{\mu(E_i)}{\mu(E_i) + \mu(E_e)\frac{\sin(\alpha)}{\sin(\beta)}} \left(1 - \exp\left\{-t\left(\frac{\mu(E_i)}{\sin(\alpha)} + \frac{\mu(E_e)}{\sin(\beta)}\right)\right\}\right), \#(1)$$

where μ is the attenuation coefficient, as measured via XANES, $I_o(E_i)$ is the intensity distribution of photons incident on the sample, $\Omega$ is the solid angle of the detector, and $\epsilon(E_e|E_i)$ is the ideal emission spectrum, representing the probability that an emission energy, $E_e$, is measured given an incident photon of energy, $E_i$. However, for nonresonant excitation, the emission spectrum is independent of the incident photon energy, *i.e.* $\epsilon(E_e|E_i) = \epsilon(E_e)$, allowing us to invert Eq. 1. to obtain an absorption-corrected spectrum,

$$\epsilon(E_e) \propto \frac{I(E_e)}{I_o(E_i)} \frac{\mu(E_i)\sin(\beta) + \mu(E_e)\sin(\alpha)}{\mu(E_i)\sin(\beta)} \left(1 - \exp\left\{-t\left(\frac{\mu(E_i)}{\sin(\alpha)} + \frac{\mu(E_e)}{\sin(\beta)}\right)\right\}\right)^{-1}. \qquad (2)$$

The right-hand side may then be numerically integrated with respect to $E_i$ across the range of incident photon energies. We found the result to be insensitive to integration bounds and consequently integrated from 8310 eV to 8370 eV for convenience.

### D. Final Experimental Parameters

Following the above strategies, we collected XANES, XES, and hybrid-spectrum measurements from a 6-μm thick foil of Ni metal acquired from EXAFS Materials. The operating parameters of the x-ray tube source were 40 kV and 200 μA. The overwhelming majority of incident photons that excite Ni 1s electrons, including both the fluorescence lines from the Au anode and the relevant part of the bremsstrahlung spectrum, are high enough in energy that our results are overwhelmingly in the isothermal limit. A manual rotation stage was integrated into the sample mount design to allow us to control the independent variables α and β or, equivalently, the effective thickness of the sample.

XES spectra were collected at sample angles α of 44 deg and 64 deg. XAFS and hybrid measurements were conducted per the procedure outlined in Section II.B. All spectra were collected in 0.25-eV increments. Multiple scans for each category of measurement were summed to provide total integration times of: 150 s per step for measurement of the incident intensity (no sample, XANES configuration); 930 s and 310 s per step, respectively, for valence XES with α = 44 deg and α = 64 deg; 620 s per step for the hybrid configuration; and 930 s per energy step for the XAFS configuration. To reduce overall measurement times, detailed valence XES scans spanned a range in energy from 8310 to 8380 eV and were later normalized to a single XES spectrum covering the full energy range from 8240 to 8380 eV.

### E. Determination of Branch Ratios

Phenomenological fits were computed using the Monte-Carlo routines available in the *BlueprintXAS* software package[56, 57]. Pseudo-Voigt functions were used to fit the $K\beta_{1,3}$, $K\beta_{2,5}$, and multi-electron spectra. Radiative Auger emission was accounted for by including an additional function as described by Enkisch *et al*[58]. The $K\beta_{1,3}$, which lies below the Ni absorption edge and thus does not require a self-attenuation correction, was used across measurements to preserve the overall intensity scale. Areas of the multi-electron peaks were then calculated and compared with relevant diagram lines to determine branching ratios. Estimating uncertainties in the branching ratio involved approximating the pseudo-Voigt integral per a method described by Lenz and Ayres[59].

## III. Results and Discussion

The spectra collected via the energy alignment procedure outlined in Section II.B. are shown in Fig. 3. The resultant difference spectrum, which shares an energy scale with our XANES measurements, provides us with an emission peak with which we can align our XES spectra – thus ensuring a common energy scale across all our measurements. Here, we find that a one-time energy shift of ~8 eV is necessary. In Fig. 4, we show the Ni XES spectra measured at two different sample rotation angles together with the same spectra after correction for sample absorption effects using the method outlined in Section II.C., this results in as much as a factor-of-two correction to the measured spectra intensity above the Fermi level. The good agreement between the corrected results at the two different sample angles confirms the validity of our treatment of sample absorption effects.

The results in Fig. 4 show several clear satellites in the spectrum above the $K\beta_{2,5}$. We identify these peaks with the Z+1 model. The approach is an established tool for the treatment of multi-electron features in arenas such as L-edge EXAFS in rare-earth minerals[60], two- and three-electron excitations in Kr XANES[61-63], and emission spectroscopy of transition metals[20, 44, 58, 64]. Despite documented shortcomings[65], many of which it shares with multiplet calculations[8, 62], this approach typically predicts accurate values of excitation thresholds and emission satellites. Specifically, satellite energies are calculated with

$$E_{\gamma\prime} = E_\gamma + BE_{Z+1} - BE_Z, \tag{3}$$

where $E_{\gamma\prime}$ is the energy of the satellite, $E_\gamma$ is the energy of the diagram line, and $BE_{Z+1}$ and $BE_Z$ are the binding energies of the electrons emitted to form spectator holes in the fully screened Z and Z+1 systems, respectively. This process yields excellent agreement between the locations of satellites predicted by the Z+1 model and the peak locations in Fig. 4-5. Having identified the various DI features, we then fit the corrected spectrum using the method described in Section II.E. We show the consequent fit in Fig. 5. A comparison between predicted and measured satellite positions is presented in Table I where a method, motivated by the convention of Druyvesteyn[4], was adopted for the [1snp] satellite by calculating the weighted average of the $np_{1/2}$ and $np_{3/2}$ binding energies according to population. The extracted branch ratios are presented, and compared with past experimental and theoretical results in Table II. These literature values were reported as probabilities, which we converted to branching ratios by dividing the satellite probability by unity less the shake probability.

While few theoretical studies are as comprehensive as that of Mukoyama *et al*[17], there exist several additional theoretical and experimental measurements with which to compare our results. Our reported value of 23% for the branching ratio of the [1s3d] satellite is in good agreement with the work of Ito *et al*[6], but not with that of Mukoyama *et al*. Nonetheless, this has been similarly observed by other authors, who have reported analogous findings in studies of Cu[6, 20, 23], Ti[25], and Sc[21], suggesting a systematic underestimation in that particular study due to an incomplete treatment off the SO process. Despite the lack of agreement with Mukoyama's predictions, our reported values agree well with the theoretical work of Lowe *et al*. While both authors' calculations were atomic in nature, Lowe employed a multi-configurational framework that was inaccessible to the earlier, single-configurational calculation of Mukoyama, but is necessary for modeling complex, open shell atoms. Furthermore, our measured branch ratio of the [1s2s] satellite is smaller than predicted by Mukoyama and the [1s3s] satellite is not present in our spectra. These observations can be explained by a suppression of the satellites by fast Coster Kronig transitions[21, 43, 48, 66]. Finally, the branching ratio of the [1s2p] we reported disagrees with the result of Mukoyama *et al* for the reasons previously discussed, but also disagree with the result of Kawatsura *et al*[43]. The latter study fit the intensity evolution of the satellite feature with the Thomas model and extracted the excitation probability from the corresponding fit parameter. The lack of agreement is then explained by the authors' own assertion that the Thomas model does account well for the intensity evolution of SO from the 2p shell.

### III. Conclusion

In conclusion, we report measurement of the x-ray emission spectrum for Ni in the regime near the Fermi level that includes both single-excitation, valence-to-core x-ray fluorescence and significant contributions from double-ionization. We have demonstrated a procedure for aligning to a universal energy scale and correcting for self-attenuation effects that is crucial when measuring features that lie beyond the single-particle Fermi level. Reported satellite positions are in good agreement with those predicted by the Z+1 approximation, and branching ratios agree well with prior experimental work for those satellites that have been previously reported. Errors associated with the branching ratios presented here are strongly influenced by the difficulty of fitting a spectrum to substantially overlapping peaks, specifically the [1s3d], which is nearly enveloped by the $K\beta_{2,5}$. Reliable, precise theoretical estimates of the satellite position and widths would allow the fit to be further constrained and lower the reported errors of the branching ratios. While the theoretical treatments discussed in this work are atomic in nature, other authors have suggested a suppression of DI features due to charge transfer effects and an influence of speciation on the weight of contributing configurations and thus multiplet structure[67]. Future studies of the valence-to-core and DI region of various Ni compounds are likely warranted, and should benefit from the methodologies demonstrated here.

**Acknowledgements**

This work was supported by the United States Department of Energy, Office of Basic Energy Sciences, under grant DE-SC0002194 and also by the Office of Science, Fusion Energy Sciences, under grant DE-SC0016251. JJK was supported by DOE grant DE-FG02-97ER45623.

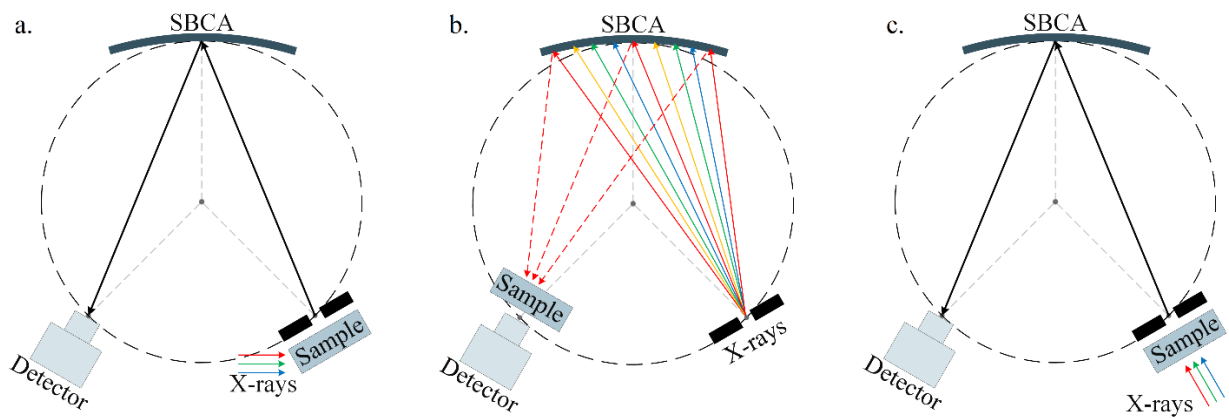

**Figure 1 | Sample Geometry.** Above are standard Rowland circle geometries used in (a) x-ray emission and (b) x-ray absorption fine structure measurements. An intermediate geometry, (c), is used to establish a common energy scale across measurements, which is necessary to correct for sample absorptive effects in XES measurements.

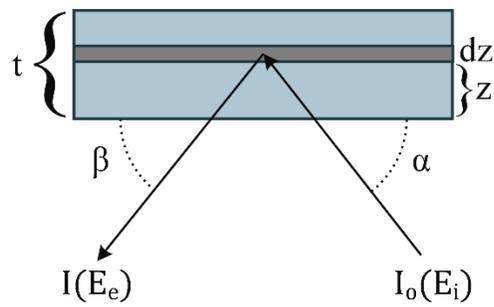

**Figure 2 | Experimental Diagrams for Energy Scale Reproduction.** X-ray photons of energy $E_i$ from a source spectrum of intensity $I_o(E_i)$ are incident at an angle α relative to the face of the sample of thickness $t$. A detector, placed at an angle β from the sample's face, measures an emission spectrum $I(E_e)$.

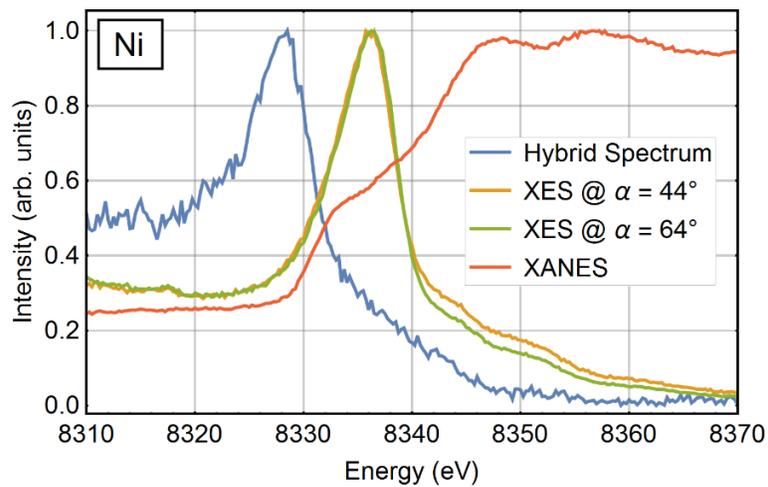

**Figure 3 | Measured Spectra from Various Sample Geometries.** Spectra from each of the three experimental configurations of Fig. 1 with the Hybrid and XANES spectra energy corrected. An energy shift of 8.75 eV aligned the Hybrid and XANES spectra to the energy scale established at the synchrotron. A comparable energy shift was also needed to align the XES data to the new, common energy scale.

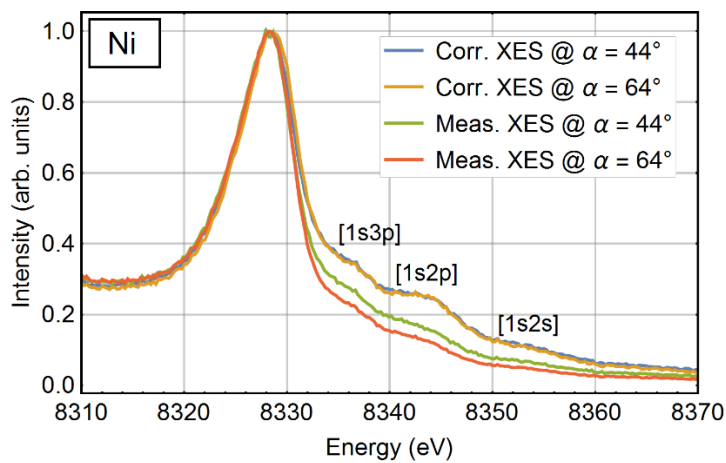

**Figure 4 | Absorption-Corrected Nickel Valence Emission.** This figure shows both the uncorrected and corrected spectra of Ni valence emission. Obtaining the correct intensity of multi-electron peaks, identified via the Z+1 approximation, is critical for theory comparison.

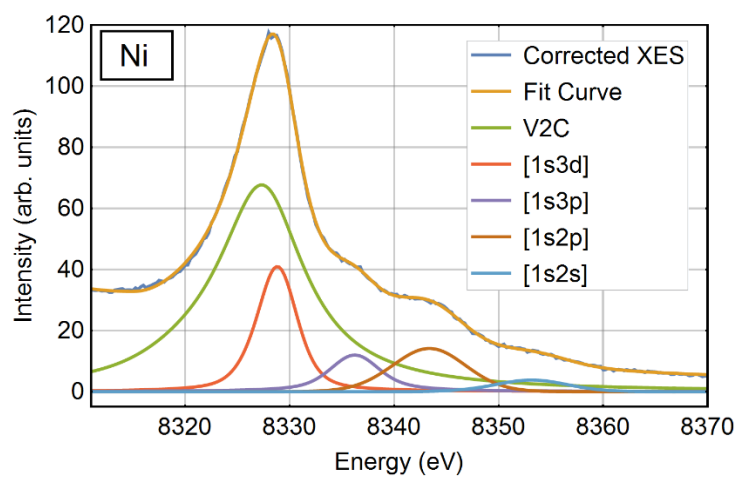

**Figure 5 | Phenomenological Fits to Ni Multi-Electron Peaks.** This figure shows the multiple pseudo-Voigt functions that were used to fit to each of the multi-electron emission peaks. These peaks' areas were used to determine the branching ratios given in Table 2.

**Table I.** Comparison between measured satellite energies and values predicted by the Z+1 model, referenced to the energy of the VTC diagram line.

| Multi-electron Transition | Z+1 (eV) | Observed (eV) |
|---|---|---|
| [1s3p] | 9.0 | 9.1 ± 0.5 |
| [1s2p] | 16.9 | 16.3 ± 0.5 |
| [1s2s] | 24.2 | 26.0 ± 0.5 |

**Table II.** Comparison between experimental and predicted branching ratios (%) of the identified DI peaks and their corresponding diagram line intensity. Literature values were converted from probabilities to branching ratios following the procedure outlined in Section III.

| DI Transition | Diagram Line | Mukoyama[17] (theory) | Hölzer[68] (exp.) | Ito[6] (exp.) | Lowe[24] (theory) | Kawatsura[43] (exp.) | Measured |
|---|---|---|---|---|---|---|---|
| [1s3p] | $K\beta_{2,5}$ | 3.19 | - | - | - | - | 9 ± 5 |
| [1s2p] | $K\beta_{1,3}$ | 0.60 | - | - | - | 0.62 | 0.15 ± 0.05 |
| [1s2s] | $K\beta_{1,3}$ | 0.12 | - | - | - | - | 0.041 ± 0.016 |
| [1s3d] | $K\beta_{2,5}$ | 11.26 | 35 | 27 | 28 | - | 23. ± 10. |